\documentclass[aps,prl,reprint,superscriptaddress,showpacs,citeautoscript]{revtex4-1}

\usepackage[pdftex]{graphicx}
\usepackage{color}

\usepackage{amsmath}
\usepackage{accents}

\makeatletter
\def\widebar{\accentset{{\cc@style\underline{\mskip10mu}}}}
\makeatother

\begin{document}

\title{Glasslike phonon excitation caused by ferroelectric structural instability}


\author{Y. Ishii}
\email{ishii@mtr.osakafu-u.ac.jp}
\affiliation{Department of Materials Science, Osaka Prefecture University, Sakai, Osaka 599-8531, Japan.}

\author{A. Yamamoto}
\affiliation{Department of Materials Science, Osaka Prefecture University, Sakai, Osaka 599-8531, Japan.}

\author{N. Sato}
\affiliation{International Center for Materials Nanoarchitectonics (WPI-MANA), National Institute for Materials Science (NIMS), Tsukuba, Ibaraki 305-0044, Japan.}
\affiliation{International Center for Young Scientists, NIMS, Tsukuba, Ibaraki 305-0047, Japan.}

\author{Y. Nambu}
\affiliation{Institute for Materials Research, Tohoku University, Sendai, Miyagi 980-8577, Japan.}
\affiliation{FOREST, Japan Science and Technology Agency, Kawaguchi, Saitama 332-0012, Japan}

\author{S. Ohira-Kawamura}
\affiliation{Materials and Life Science Division, J-PARC Center, Tokai, Ibaraki 319-1195, Japan.}

\author{N. Murai}
\affiliation{Materials and Life Science Division, J-PARC Center, Tokai, Ibaraki 319-1195, Japan.}

\author{T. Mori}
\affiliation{International Center for Materials Nanoarchitectonics (WPI-MANA), National Institute for Materials Science (NIMS), Tsukuba, Ibaraki 305-0044, Japan.}

\author{S. Mori}
\affiliation{Department of Materials Science, Osaka Prefecture University, Sakai, Osaka 599-8531, Japan.}



\begin{abstract}

Quest for new states of matter near an ordered phase is a promising route for making modern physics forward.
By probing thermal properties of a ferroelectric (FE) crystal Ba$_{1-x}$Sr$_x$Al$_2$O$_4$, we have clarified that low-energy excitation of acoustic phonons is remarkably enhanced with critical behavior at the border of the FE phase.
The phonon spectrum is significantly damped toward the FE phase boundary and transforms into glasslike phonon excitation which is reminiscent of a boson peak.
This system thus links long-standing issues of amorphous solids and structural instability in crystals to pave the way to controlling lattice fluctuation as a new tuning parameter.

\end{abstract}


\maketitle

Amorphous solids are generally known to exhibit low-temperature anomalies in thermal properties: a $T$-linear term in low-temperature heat capacity ($C$), a considerable enhancement in $C/T^3$, and a plateau in thermal conductivity \cite{GlassCp_PRB1971,Glass-Cp_PRL,MERTIG1984369,QueenPRL, Buchenau_PRB, BosonPeak_SSC, BosonPeak_Nakayama,amorphous_kappa1,amorphous_kappa2,Slack}. 
Phonon picture is no longer appropriate in the amorphous systems because of a lack of periodic structure.
Early works on the vibrational spectra have revealed that a single broad maximum called a boson peak exists approximately at 4 meV, below which excitations of local atomic vibration are increased \cite{Buchenau_PRL}. 

Despite a large number of continuous efforts \cite{BP_Localized,BosonPeak_vanHoveSingularity,LocalFavoredStructure,MolecularFragments,AcousticMode_PRL1,AcousticMode_PRL2,AcousticMode_NatMater}, these anomalous features have long been one of the most puzzling issues in modern physics for almost half century.
What is more enigmatic is a number of reports on crystals that show the glasslike thermal properties.
These crystals commonly include disorders, such as molecular disorders, minimal disorders, or orientational disorders \cite{molecular_crystal,MinimalDisorder,OrientationalGlass,KBr-KCN,TMori}. 
Crystals with lone pairs or strong anharmonicity are also examples\cite{lonepairs,I-clathrate,tetrahedrite}.
Several attempts have recently been made to connect the low-energy excitation in amorphous solids and phonon dispersion. 
According to a study on the vibrational density of states (DOS) of amorphous solids, 
the boson peak originates from piling up of the acoustic states near the pseudo-Brillouin-zone boundary \cite{Chumakov_PRL}.
Recent theoretical work has pointed out 
that the boson peak can be understood in the sense of elastic mode propagation and viscous damping by disorders in crystals \cite{Baggioli_PRL}.
For a relatively high-temperature region above $\sim$50 K, the unified theory that can deal both with crystals and amorphous solids has been proposed \cite{Simoncelli}.
It is thus highly desired to establish the connection between the atomic vibration in amorphous solids and the phonon modes in crystals.

The nature of phonons modified by disorders is also the emerging subject from a more condensed matter perspective, such as a potential of coupling between disorders and itinerant electrons \cite{Coak}.
A striking work has been provided by C. Setty \cite{Setty}, who has theoretically demonstrated that superconducting $T_{\rm c}$ is enhanced when weakly dissipated bosons act coherently.
In fact, there have been several reports on disorder-enhanced $T_{\rm c}$, as exemplified by the proton-irradiated La$_{1.875}$Ba$_{0.125}$CuO$_4$ \cite{LaBaCuO4}, electron-irradiated FeSe \cite{irradiatedFeSe}, and elemental superconductivity in amorphous substances. 
To extend this idea of disorders to ferroelectric lattice instability is particularly interesting, as mentioned by M.N. Gastiasoro {\it et al.} \cite{Gastiasoro1}.
Indeed, a recent study \cite{Gastiasoro2} has revealed the local $T_{\rm c}$ enhancement due to dislocation-induced strain in the doped SrTiO$_3$, which has attracted increasing interest in terms of superconductivity mediated by ferroelectric fluctuation \cite{Rowley_NatPhys,Edge_PRL115,Ferroelectric_Super,Kozii}.
In order to gain a universal understanding of the mysterious thermal properties in amorphous solids and use it to explore novel states of matter, 
it is of central importance to clarify how the phonon spectra are modified by disorders in crystals especially in the low-energy region.

This study aims to show how the ferroelectric lattice instability affects phonon excitation by focusing on an improper ferroelectric crystal, BaAl$_2$O$_4$.
This compound has a crystal structure that comprises an AlO$_4$-tetrahedral network with six-membered cavities occupied by Ba ions.
The high-temperature phase of BaAl$_2$O$_4$ (a space group $P6_322$) possesses a soft mode, which is a transverse acoustic (TA) mode characterized as M$_2$ in irreducible representation.
It condenses at $T_{\rm C}$ = 450 K, resulting in the second-order type ferroelectric (FE) transition \cite{Ishii_PRB93}. 
The Ba site can be replaced by Sr ions in all proportions.
As shown in the phase diagram of Ba$_{1-x}$Sr$_x$Al$_2$O$_4$ [Fig. 1(a)], the FE transition is rapidly suppressed by a small amount of Sr substitution.
Outside the FE phase, the FE transition is not observed down to 2 K \cite{Ishii_PRB94}.

This compound has another soft mode characterized as the K$_2$ mode, of which energy competes with the ferroelectric M$_2$ mode.
The K$_2$ mode temporarily forms another ordered structure ($P6_3$ ($\sqrt{3}a$)) outside the ferroelectric phase in a narrow compositional window of $x=0.07$--0.15. 
This temporal structural order is also rapidly suppressed and disappears at $x\sim0.15$.  
Further increase in the Sr composition leads to the first-order phase transition at $x\sim 0.6$.
In addition, the both boundaries at $x=0.07$ and 0.6 are ambiguous, as indicated in Fig. 1(a) by broken lines \cite{Ishii_SciRep,Ishii_PRB94,Ishii_SSC249}.
A similar ambiguous phase boundary has also been reported for Mo$_{1-x}$Nb$_x$Te$_2$ \cite{MoTe2}, which is known as a polar metal.
Fig. 1(b) describes the unit cell relation among the various superstructures of Ba$_{1-x}$Sr$_x$Al$_2$O$_4$.

In the middle compositional window of $x=0.15$--0.5, the crystal shows none of the superstructures but, instead, shows thermal diffuse scattering, which presumably arises from the M$_2$ and K$_2$ modes \cite{Ishii_SciRep,Ishii_PRB93}.
Interestingly, the intensity of the thermal diffuse scattering strongly depends on temperature and exhibits maximum at $T^*$, although the dynamic character is unclear.
Recently, we have found that Ba$_{1-x}$Sr$_x$Al$_2$O$_4$ in this compositional window exhibits a $T$-linear term at low temperature and a large hump in $C/T^3$ curves \cite{Ishii_PRM}, both of which are the typical characters of amorphous solids. 

As described above, this compound is a suitable system to investigate how ferroelectric fluctuations affect the dynamic properties of crystals.
In this Letter, we demonstrate that lattice dynamics are considerably modified toward the border of the FE phase.
While overall excitations of phonon modes are entirely damped, low-energy excitations of the acoustic modes near the $\Gamma$ point are significantly enhanced.
Our findings provide new insights into not only the classical but modern enigma observed in thermal properties of amorphous and crystalline solids but also the exotic condensed states governed by structural fluctuation, which should accelerate crosscutting researches to pioneer a new state of condensed matter where both natures of crystalline and amorphous solids are precisely tuned and mutually interact.

\begin{figure}[t]
\begin{center}
\includegraphics[width=85mm]{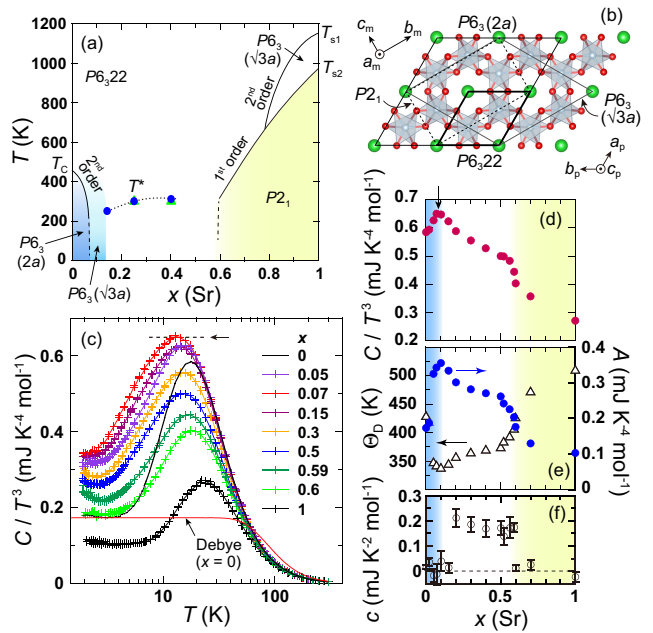}
\caption{\label{PhaseDiagram} 
(a) Structural phase diagram of Ba$_{1-x}$Sr$_x$Al$_2$O$_4$ together with the space group of each phase. The phase boundaries were determined by x-ray diffraction and dielectric constant measurements \cite{Ishii_SciRep,Ishii_PRB94,Kawaguchi_PRB,Ishii_SSC249,Supple}. 
(b) Various cell settings of Ba$_{1-x}$Sr$_x$Al$_2$O$_4$.
(c) $C/T^3$ as a function of temperature. Debye contribution for $x=0$ is depicted by a red line. The value of $C/T^3$ peak is defined by a broken line.
The right panels show the $x$ dependence of (d) the value of $C/T^3$ peak, (e) Debye temperature, $\Theta_{\rm D}$ (left axis), together with the $T^3$ coefficient, $A$ (right axis), and (f) $T$-linear coefficient, $c$. 
$\Theta_{\rm D}$ and $c$ are evaluated from the slope and intercept of the linear fit by $C/T=AT^2+c$, respectively.  
}
\end{center}
\end{figure}

Heat capacity and thermal conductivity were measured for the Ba$_{1-x}$Sr$_x$Al$_2$O$_4$ polycrystalline samples of $x=0$--1 using the heat-relaxation and steady-state methods, respectively, in a physical property measurement system (PPMS, Quantum Design).
Phonon calculations were performed by using PHONOPY \cite{phonopy,phonopy2} in conjunction with VASP code \cite{vasp}.
Sample preparation methods are described in detail \cite{Supple}.
For the inelastic neutron scattering (INS) experiment, the powder samples of $x$ = 0, 0.03, 0.07, 0.2, 0.3 (mass $\sim$5 g each) were separately enclosed in thin-walled aluminum cans with He exchange gas. 
The INS measurement was performed for these samples on BL14 AMATERAS at J-PARC \cite{AMATERAS}. The incident neutron energies ($E_{\rm i}$) were set at $E_{\rm i}$ = 52.45, 17.255, and 8.485 meV. 
A top-loading closed cycle refrigerator was used for temperature control. Obtained data were visualized using Mslice \cite{DAVE} after data reduction \cite{Utsusemi}. 
Phonon spectra were created using OCLIMAX \cite{OCLIMAX}.

Temperature variation of the $C/T^3$ of $x$ = 0--1 is shown in Fig. 1(c) together with Debye contribution of $x=0$ in red.
As expected from the Debye model for three dimensional insulating crystals, the $x=0$ sample shows an almost constant value in the lowest-temperature range.
In the middle-temperature range, it shows a prominent peak approximately at 20 K.
The deviation from the Debye contribution comes from acoustic modes with non-linear dispersion near the zone boundary and optical modes.
As evidenced by the $C/T^3$ value which increases as $x=0.07$ is approached, low-energy excitation of phonon is enhanced near the boundary composition of the FE phase.
The maximum value of $C/T^3$, indicated by an arrow in Fig. 1(c), shows a clear anomaly at $x=0.07$, as shown in Fig. 1(d).
The $T^3$ coefficient, $A$, varies with $x$ in a similar fashion to the maximum value of $C/T^3$, and correspondingly, $\Theta_{\rm D}$ shows a dip at that composition, as shown in Fig. 1(e). 
The $\Theta_{\rm D}$ is evaluated from Fig. S2 \cite{Supple}, where the experimental data are fitted linearly by using $C/T = AT^2 + c$.
Figure 1 (f) shows the $T$-linear coefficient, $c$, as a function of $x$.
One can find that the $T$-linear coefficient suddenly shows non-zero value in the middle compositional window, $x=0.2$--0.59, where none of the low-temperature ordered structures, $P6_3$ ($2a$), $P6_3$($\sqrt{3}a$) and $P2_1$, are observed.
This non-zero $c$ value verifies the glasslike nature in this compositional window. 
At the other compositions, the $T$-linear coefficient yields zero value within an experimental error.
The large hump in $C/T^3$ is also the common feature in amorphous solids.

\begin{figure}[t]
\begin{center}
\includegraphics[width=86mm]{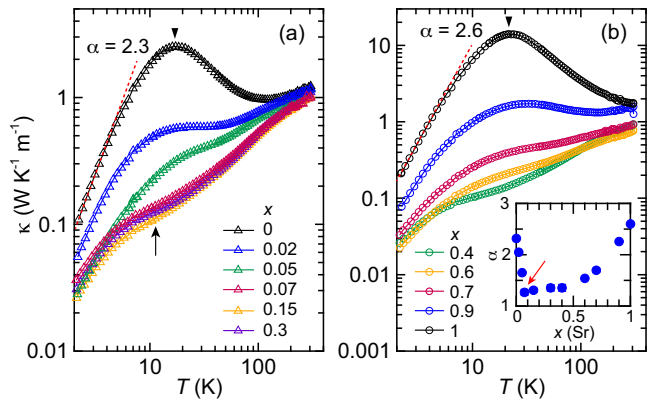}
\caption{\label{kappa} 
Log-log plots of the thermal conductivity as a function of temperature for Ba$_{1-x}$Sr$_x$Al$_2$O$_4$ of (a) $x$ = 0--0.3 and (b) $x$ = 0.4--1. 
Error bars are smaller than the symbol size.
Red broken lines show the power fit using $\kappa \propto T^{\alpha}$ for the experimental data of $x=0$ and $x=1$. A plateau is observed for $x$ = 0.07--0.4 near 10 K as indicated by a black arrow.
Inset shows the power exponent $\alpha$ plotted against $x$.   
}
\end{center}
\end{figure}

\begin{figure}[t]
\begin{center}
\includegraphics[width=85mm]{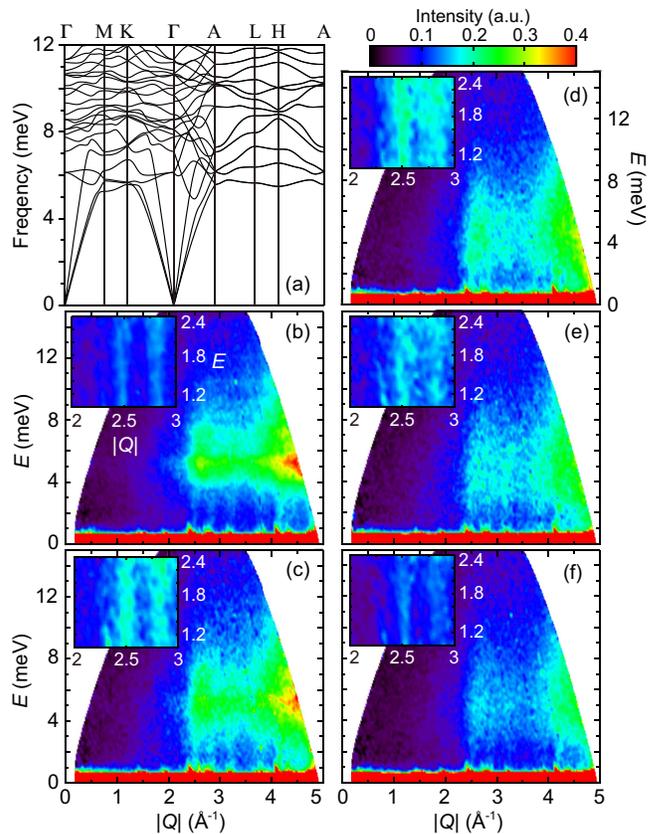}
\caption{\label{INS_SQw} 
(a) Phonon-dispersion relations calculated on the ferroelectric $P6_3$ ($2a$) structure of BaAl$_2$O$_4$. The calculation uses a $2\times2\times3$ supercell of the unit cell.
Dynamical structure factor, $S$($Q$, $E$), for Ba$_{1-x}$Sr$_x$Al$_2$O$_4$ powder samples of (b) $x=0$, (c) 0.03, (d) 0.07, (e) 0.2, and (f) 0.3 measured at 100 K with $E_{\rm i} = 17.255$ meV. 
The insets show the enlarged view in the range of $E=1.0\sim2.4$ meV and $|Q| = 2\sim3$ \AA \; obtained with $E_{\rm i} = 8.485$ meV.
Al-sample can background has been subtracted in each scan.
}
\end{center}
\end{figure}

\begin{figure}[t]
\begin{center}
\includegraphics[width=85mm]{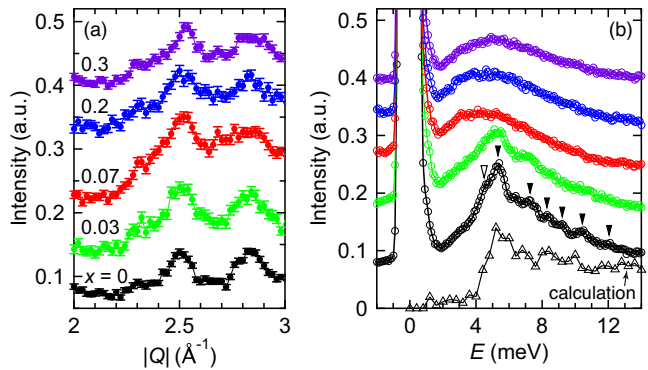}
\caption{\label{QEcut} 
(a) Energy-integrated inelastic spectra over the energy transfer $E=1\sim2.3$ meV with $E_{\rm i} = 8.485$ meV. For clarity, profiles are offset by 0.05 and 0.1 for $x=0.03$ and other compositions, respectively. (b) $Q$-integrated inelastic spectra over $|Q|=2\sim4$ \AA$^{-1}$ with $E_{\rm i} = 17.255$ meV. Profiles are offset by 0.08 for clarity.
}
\end{center}
\end{figure}

The glasslike nature is also confirmed in the thermal conductivity.
Figures 2(a) and (b) show the temperature dependence of $\kappa$.
In crystals, it generally behaves as $\kappa \propto T^{\alpha} (\alpha=3)$ at the lowest temperatures, followed by $\kappa \propto \exp(E/T)$ at several 10 K, and accordingly shows a peak around 20 K.
This characteristic peak is clearly observed at the end members, as indicated by solid triangles in Figs. 2(a) and (b).
The low-temperature exponents $\alpha$ for $x=0$ and 1 yield 2.3 and 2.6, respectively.
The smaller values than the ideal $\alpha=3$ are explained by the isotope effect.
The peak observed in $x=0$ is suppressed as the boundary composition $x=0.07$ is approached and entirely disappears outside the FE phase.
Instead, the plateau-like region appears at $x=0.07$--0.4 as indicated by a black arrow [Fig. 2(a)].
The large peak resurges as $x$ increases further.
The inset of Fig. 2(b) summarizes the exponents $\alpha$. 
Just as the $\Theta_{\rm D}$, it shows a dip at $x=0.07$, as indicated by a red arrow.
On the other hand, there is no anomaly at $x=0.6$, which is the border of the first-order structural phase transition.

It is curious that the $x=0$ sample shows an increasing trend above $\sim$100 K on heating, which is contrary to the general expectation of high-temperature behavior, $\kappa \propto T^{-x} (x=1 \sim 2)$.
Deviation from the $T^{-x}$ dependence has been theoretically predicted in the system with a strong anharmonicity \cite{Tadano}.
Strong thermal diffuse scattering observed for $x=0$ in this temperature region \cite{Ishii_PRB93} should be responsible for this departure.

Figure 3(b) shows the inelastic neutron scattering spectrum of $x = 0$ obtained at 100 K with the incident energy $E_{\rm i} = 17.255$ meV. 
The calculated phonon dispersion for the $P6_3$ phase of $x=0$ and the simulated INS spectrum based on this calculation are shown in Fig. 3(a) and Fig. S3 \cite{Supple}, respectively.
Overall, the simulation well reproduces the experimental result.
The prominent excitations observed in Fig. 3(a) at $\sim$5 meV mainly come from the optical modes that have a flat dispersion. 
Weak streaks observed at 2.5, 2.8, 3.2, 3.8, and 4.2 \AA$^{-1}$ are ascribed to the dispersive acoustic modes.
Figures 3(c)--(f) show the INS spectra of $x=0.03$, 0.07, 0.2, and 0.3. 
Interestingly, the energy distribution of the scattering intensity deforms and shifts toward lower-energy region as $x$ increases.
As a result, the scattering intensity at the low-energy region below $\sim$4 meV is increased, as $x=0.07$ is approached.

The insets of Figs. 3(b)--(f) show the spectra of the low-energy region measured with $E_{\rm i} = 8.485$ meV.
Each spectrum is integrated over the energy transfer $E=1\sim2.3$ meV, as shown in Fig. 4(a).
The $x=0$ sample shows two peaks at $Q=2.5$ and 2.8 \AA$^{-1}$, which $Q$ values mainly correspond to the 111 and 112 Bragg reflections, respectively.
These excitations strongly depend on $x$, showing a maximum at the boundary composition, $x=0.07$, as we observed in the measurements with $E_{\rm i}=17.255$ meV.
This increased low-energy excitation of the acoustic modes should unambiguously be the leading cause for the enhanced low-temperature heat capacity.
After showing a maximum at $x=0.07$, the low-energy excitation decreases as $x$ increases.

To see the energy distribution, the INS intensity measured with $E_{\rm i}=17.255$ meV and the calculated one were integrated over $|Q| = 2\sim4$ \AA$^{-1}$, as shown in Fig. 4 (b). 
The calculation well reproduces the overall feature of the experimental result for $x$ = 0.
As indicated by solid triangles, several sharp peaks are observed at $x$ = 0, which are from overlapping of several optical modes.
In addition, the most prominent peak has a shoulder, as indicated by an open triangle, which is ascribed to the acoustic modes near the Brillouin zone boundary.
The sharp peaks are suppressed at $x=0.03$.
Surprisingly, they entirely disappear at $x=0.07$, and the spectrum shows a broad feature that is reminiscent of a boson peak.
This fact directly indicates that the ferroelectric instability with the acoustic character damps not only the acoustic excitations but also the optical excitations.
The glasslike thermal nature of Ba$_{1-x}$Sr$_x$Al$_2$O$_4$ crystals is governed by this damped phonon spectrum.
The enhanced intensity in the low-energy region below $\sim$4 meV evidences that the excitation of long-wavelength acoustic modes near the $\Gamma$ point is enhanced toward $x=0.07$.

In other words, the broad feature of the INS spectrum testifies to the fact that the suppression of the FE phase causes the structural disorder.
This observation can lead to the following interpretation;
suppression of the ferroelectric order induces atomic fluctuations where the original instability of the soft modes resides as correlated disorders. 
These correlated disorders are probably the static ones.
They significantly prohibit the propagation not only of the acoustic modes but also of the optical modes and, therefore, damp the overall excitations.

The $Q$-integrated spectra shown in Fig. 4(a) authenticate the piling-up picture proposed by Chumakov {\it et al}. \cite{Chumakov_PRL} and the theoretical prediction that the boson peak is the broadening and lowering of the lowest van Hove singularity \cite{AcousticMode_PRL1,AcousticMode_PRL2}.
However, as is observed in Figs. 3 and 4, 
the streaks of the acoustic modes uprising from the Bragg peaks are clearly observed even after the fine structure stemming from the phonon dispersion is entirely lost by the suppression of the ferroelectric phase.
Furthermore, the periodic $Q$ dependence is also retained in the higher-energy excitations of optical modes.
These facts indicate that the phonon picture is preserved even after the FE phase transition is entirely suppressed, which is a completely different nature from amorphous solids.

It is noteworthy that the critical behavior observed in $\Theta_{\rm D}$ and $\alpha$ may probably be a sign of the structural quantum criticality in this system.
Notably, such the behavior is absent at $x=0.6$, 
the first-order type structural phase boundary on the opposite side of the phase diagram, 
where the maximum value of $C/T^3$, the coefficient $A$, and the $\Theta_{\rm D}$ only show a kink.
An abrupt decrease in $\Theta_{\rm D}$ at a structural quantum critical point has been reported in (Sr$_{1-x}$Ca$_x$)$_3$Rh$_4$Sn$_{13}$ \cite{Sr3Rh4Sn13_PRL} and LaCu$_{6-x}$Au$_x$ \cite{LaCu6-xAux}, which are known as structural quantum materials.
Iron-based superconductor Ba(Fe$_{1-x}$Co$_x$)$_2$As$_2$ \cite{yoshizawa1,yoshizawa2} and ferroelectric superconductor (Sr,Ca)TiO$_{3-\delta}$ \cite{Ferroelectric_Super} are also accepted to show a structural quantum critical point (SQCP), although the lattice dynamics have not been thoroughly explored.
In order to establish the essential and universal understanding of the nature of SQCP, it is necessary to accumulate data on the lattice dynamics of various candidate materials.

In this study, the glasslike natures in the heat capacity and thermal transport properties have been rationalized based on the INS experiment.
Phonon modes with high energy are damped at the border of the FE phase where the soft-mode condensation is inhibited, and concurrently, the spectrum shifts toward lower energy.
As a result, the acoustic excitation near the $\Gamma$ point is remarkably increased. 
Although the phonon spectrum loses energy dependence, the $Q$ dependence is preserved due to the periodicity of the crystal lattice.
In addition, this system exhibits the critical behavior at the border of the FE phase, which indicates inherent lattice instability underlies there.
Our findings provide deep insights into the recent trends of modern physics that include viscosity of phonon propagation \cite{Baggioli_PRL}, disorder- or strain-induced superconductivity \cite{FeSe-film1,FeSe-film2}, unifying theory of amorphous solids and crystals \cite{AcousticMode_NatMater,Simoncelli,Simon}, and structural quantum criticality where exotic electron-phonon interactions should exist, all of which are definitely in the forefront of exploring new states of condensed matter.

\begin{acknowledgements}

This work was supported by a JSPS Grant-in-Aid for Scientific Research on Innovative Areas ``Mixed-anion'' (No. 16H06441, 17H05473, 17H05487, 19H04683, 19H04704), JSPS KAKENHI (No. 17H06137, 20H01844), and JST Mirai (No. JPMJMI19A1).
The INS experiments were performed at AMATERAS installed at BL14 in the Materials and Life Science Experimental Facility (MLF) in Japan Proton Accelerator Research Complex (J-PARC) (Proposal No. 2019B0022).

\end{acknowledgements}


\end{document}


\begin{center}
Supplemental Material
\end{center}

\title{Glasslike phonon excitation caused by ferroelectric structural instability}


\author{Y. Ishii}
\email{ishii@mtr.osakafu-u.ac.jp}
\affiliation{Department of Materials Science, Osaka Prefecture University, Sakai, Osaka 599-8531, Japan.}

\author{A. Yamamoto}
\affiliation{Department of Materials Science, Osaka Prefecture University, Sakai, Osaka 599-8531, Japan.}

\author{N. Sato}
\affiliation{International Center for Materials Nanoarchitectonics (WPI-MANA), National Institute for Materials Science (NIMS), Tsukuba, Ibaraki 305-0044, Japan.}
\affiliation{International Center for Young Scientists, NIMS, Tsukuba, Ibaraki 305-0047, Japan.}

\author{Y. Nambu}
\affiliation{Institute for Materials Research, Tohoku University, Sendai, Miyagi 980-8577, Japan.}
\affiliation{FOREST, Japan Science and Technology Agency, Kawaguchi, Saitama 332-0012, Japan}

\author{S. Ohira-Kawamura}
\affiliation{Materials and Life Science Division, J-PARC Center, Tokai, Ibaraki 319-1195, Japan.}

\author{N. Murai}
\affiliation{Materials and Life Science Division, J-PARC Center, Tokai, Ibaraki 319-1195, Japan.}

\author{T. Mori}
\affiliation{International Center for Materials Nanoarchitectonics (WPI-MANA), National Institute for Materials Science (NIMS), Tsukuba, Ibaraki 305-0044, Japan.}

\author{S. Mori}
\affiliation{Department of Materials Science, Osaka Prefecture University, Sakai, Osaka 599-8531, Japan.}




\maketitle

Regarding the sample preparation procedure, powder samples of Ba$_{1-x}$Sr$_x$Al$_2$O$_4$ were synthesized using a conventional solid-state reaction, as described in \cite{Ishii_SciRep}. 
For the measurements of heat capacity and dielectric constant, obtained powder was pressed under hydrostatic pressure after uniaxial press and then sintered at 1450$^{\circ}$C for 48 h.
For the thermal conductivity measurement, powder samples were densified by spark plasma sintering with a maximum temperature of 1300--1350$^{\circ}$C for 10 min to achieve sample density as high as $\approx$100\% of the theoretical value.

\vspace{10mm}

\begin{figure}[h]
 \begin{center}
   \includegraphics[width=150mm]{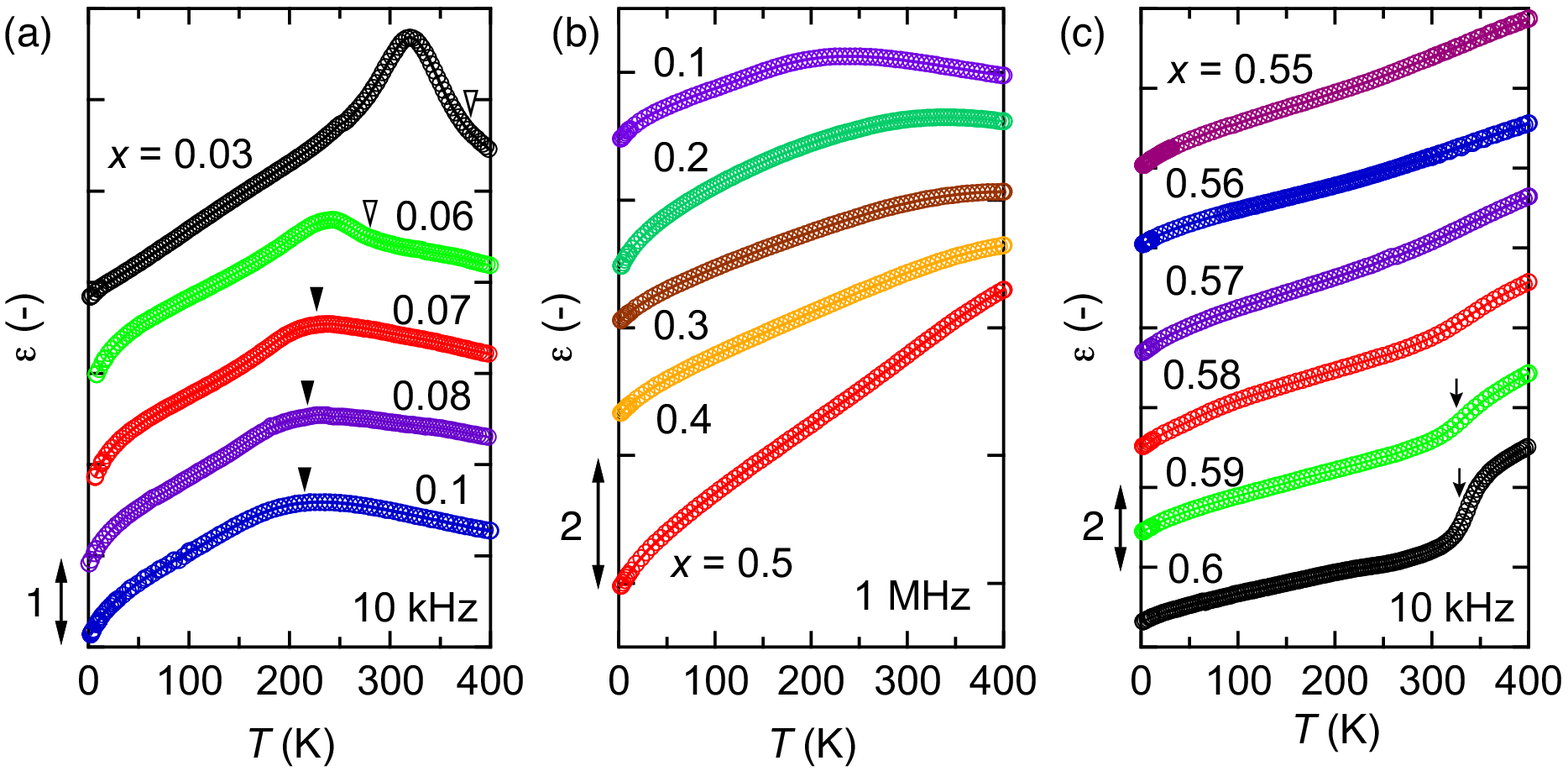}
  \caption{Temperature variation of dielectric constants of Ba$_{1-x}$Sr$_x$Al$_2$O$_4$ measured by a two probe method. (a) $x=0.03\sim0.1$ measured with 10 kHz, (b) $x=0.1\sim0.5$ measured with 1 MHz, and (c) $x=0.55\sim0.6$ measured with 10 kHz.
 A sharp signal at the ferroelectric transition temperature ($T_{\rm C}$) is suppressed as $x$ increases and transforms into a broad feature at $x>$ 0.07, as indicated by open and closed triangles, respectively.
 Above $x$ = 0.59, an anomaly due to the first-order transition is observed, as indicated by arrows.
  }
  \label{dielectric_all}
 \end{center}
\end{figure}

\clearpage

\begin{figure}[t]
 \begin{center}
   \includegraphics[width=150mm]{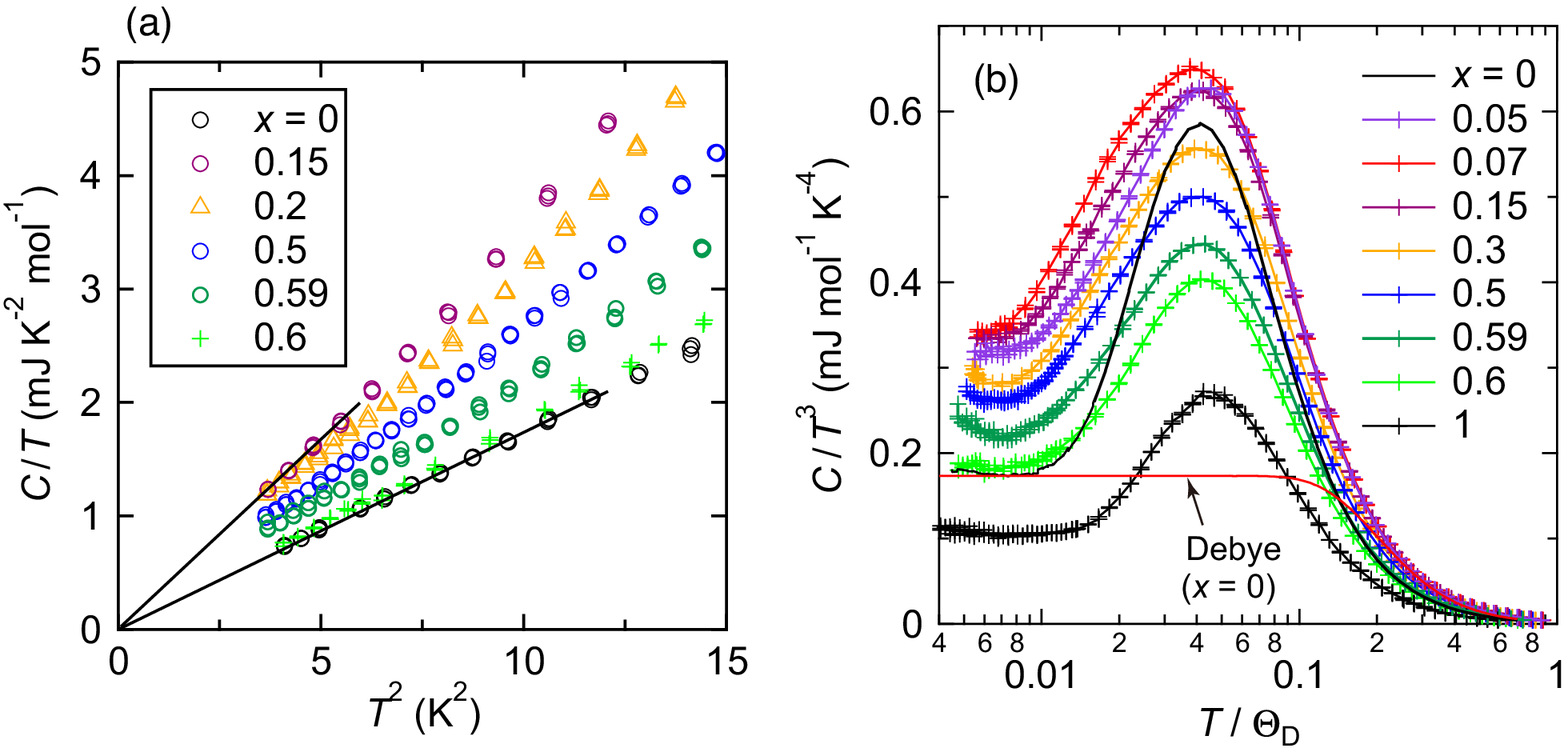}
  \caption{Heat capacity of Ba$_{1-x}$Sr$_x$Al$_2$O$_4$. (a) $C/T$ vs. $T^2$ plots. Debye temperature ($\Theta_{\rm D}$) and the $T$-linear coefficient ($c$) are evaluated from the linear fit by $C/T=AT^2 +c$. The $T$-linear term appears above $x=0.2$. (b) $C/T^3$ as a function of temperature divided by $\Theta_{\rm D}$. 
  The peak of $C/T^3$ is observed at $\approx0.04 \Theta_{\rm D}$ in all samples.}
 \end{center}
\end{figure}

\begin{figure}[b]
 \begin{center}
   \includegraphics[width=100mm]{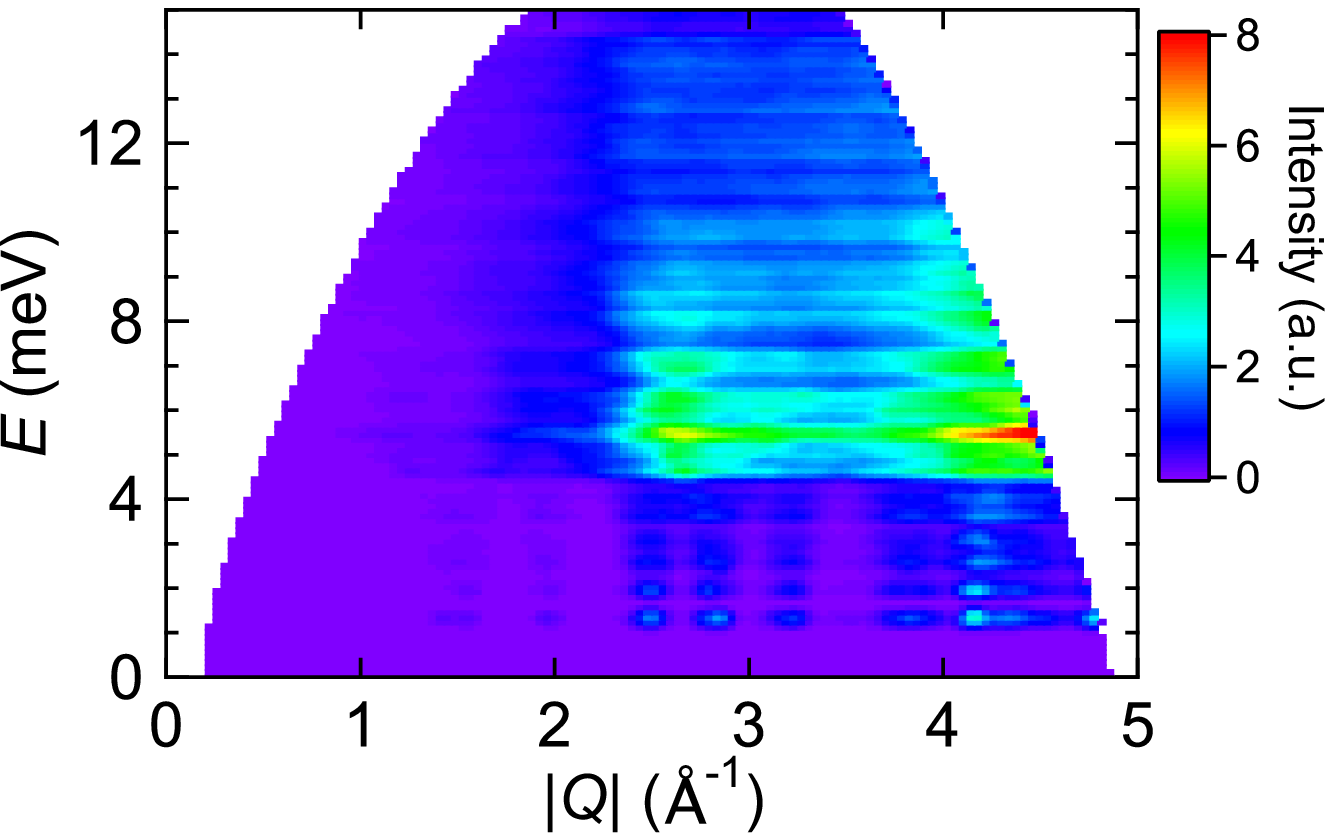}
  \caption{$S(Q,E)$ pattern of BaAl$_2$O$_4$ calculated using the OCLIMAX program \cite{OCLIMAX} based on the first-principles calculation \cite{phonopy,phonopy2,vasp}.}
 \end{center}
\end{figure}

\clearpage

\begin{figure}[t]
 \begin{center}
   \includegraphics[width=90mm]{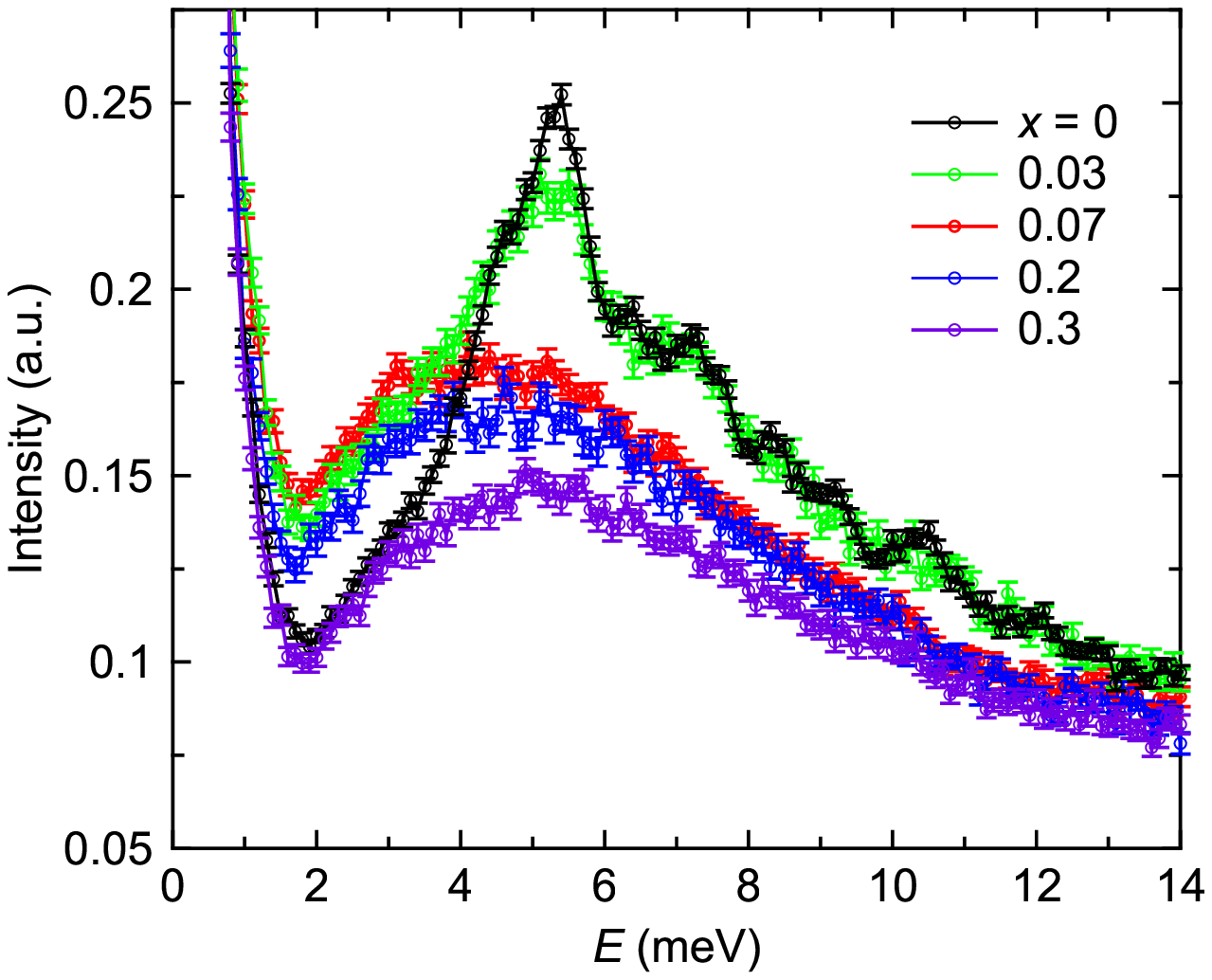}
  \caption{$Q$-integrated inelastic spectra over $|Q|=2\sim4$ \AA$^{-1}$ displayed without the vertical offset. (a) $x=0$, (b) 0.03, (c) 0.07, (d) 0.2, and (e) 0.3. Data were collected with $E_{\rm i}$ = 17.255 meV at 100 K.}
 \end{center}
\end{figure}